# THE PERFORMANCE OF THE THIN NaI(Tl) DETECTOR PICO-LON FOR DARK MATTER SEARCH

K.Harada, K.Fushimi, S.Nakayama, R.Orito, S.Iida, S.Ito,

Faculty of Integrated Arts and Science The University of Tokushima,

1-1 Minami Josanjimacho Tokushima city, Tokushima, 770-8502, JAPAN

H.Ejiri, T.Shima

Research Center for Nuclear Physics(RCNP),

10-1, Mihogaoka, Ibaraki city, Osaka, 567-0047, JAPAN

R.Hazama,

Faculty of Engineering, Hiroshima University,

1-4-1 Kagamiyama, Higashi Hiroshima city, Hiroshima, 739-8527, JAPAN

E.Matsumoto, H.Ito,

Horiba Ltd

2 Kisshoin Miyanohigashimachi, Minami-ku, Kyoto city, Kyoto, 601-8510, JAPAN

K.Imagawa

I. S. C. Lab.

3-8-11 Kitakasugaoka, Ibaraki city, 567-0048 Osaka, JAPAN

## 1. Introduction

The dark matter is of great interest in the field of particle astrophysics and nuclear physics. Components of the universe have been studied by many cosmological observations. The cosmological parameters were derived from the WMAP data combined with the distance measurements from the type Ia supernovae and the Baryon Acoustic Oscillations in the distribution of galaxies [1]. The density parameters of the baryon, the cold dark matter and the cosmological constant are derived as $\Omega_b h^2 = 0.02267^{+0.00058}_{-0.00059}$, $\Omega_c h^2 = 0.1131 \pm 0.0034$ and $\Omega_\Lambda = 0.726 \pm 0.015$, respectively., where, $h = \frac{H_0}{100}$ is Hubble constant [1]. The dark matter distribution in the Universe has been studied by means of the micro lensing observation. The three-dimensional distribution map was reported by COSMOS collaboration [2]. From many observations described above, CDM (Cold Dark Matter) is the promising candidate of the main ingredient of the dark matter. WIMPs (Weakly Interacting Massive Particles) are the most promising candidate of CDM.

The nucleus - WIMPs interaction is dominated by the weak interaction [3]. The nucleus - WIMPs interaction is classified into three types; Spin-independent elastic scattering (SI), spin-dependent elastic scattering (SD) and inelastic excitation of nucleus (EX) [4] .

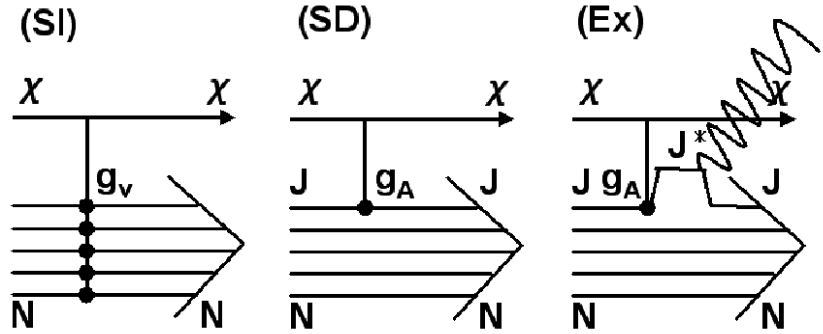

Fig.1. Schematic Feynman diagram of WIMPs-nucleus interaction. "N" and "χ" stand for the nucleus and the WIMPs, respectively. (SI) is spin independent elastic scattering. (SD) is spin dependent elastic scattering. (Ex) is inelastic scattering [4, 5].

The processes for WIMPs-nucleus interaction are schematically illustrated in Fig.1. The properties of each interaction are described below.

(i) The spin independent elastic scattering is due to coherent interactions by all the nucleon. The interaction cross section depends on the square of the mass number of target nucleus.

(ii) The spin dependent elastic scattering is due to the interaction of only one nucleon which carries the nuclear spin in case of an odd mass number nucleus. The cross section depends on the nuclear spin matrix element $\lambda^2 J(J+1)$, where $J$ is total nuclear spin and $\lambda$ is the fraction of spin which the valence nucleon carries.

(iii) Nuclear excitation takes place if the target nucleus has low lying and easily excited states. The nuclear excitation is followed by a gamma ray emission.

The highly sensitive detector PICO-LON (Planar Inorganic Crystals Observatory for Low-background Neutr(al)ino) is suitable for all the types of the interactions. In the present paper, the performance of the thin and large area NaI(Tl) scintillator developed to search for WIMPs dark matter particles. The good energy resolution and position resolution which are enough to search for WIMPs were obtained.

## 2. THE THIN NaI(Tl) DETECTOR

PICO-LON (Planar Inorganic Crystals Observatory for Low-background Neutr(al)ino) has been developed to search for the WIMPs dark matter. Many projects proposed the highly sensitive detectors using various target nuclei. NaI has good sensitivity for both the spin independent and the spin dependent elastic scatterings because the $^{127}$I isotope of NaI has both the finite spin and the large mass number *A*. We have proposed the segmentation of the detector system as a sensitive method to search for WIMPs.

The most important objective of PICO-LON is to discriminate between WIMPs and background events by means of a segmentation system. $^{127}$I is excited easily by inelastic scattering with WIMPs, accompanied by 57.6keV γ-ray if WIMPs have large kinetic energy. Large volume crystal detectors do not discriminate between the γ-ray and the recoil energy. Thin segmented NaI(Tl) has the following advantages [6, 7, 8].

(i) The range of a recoil nucleus is short enough to stop in one segment. Thus signals of both the elastic and inelastic events are observed from only one segment (Recoil energy in Fig.3).

(ii) In the case of the inelastic excitation of nucleus or the atomic electron, the low energy *γ* ray or X ray is emitted [5]. If the *γ* ray or X ray escapes from the segment and it is detected in the neighbouring segment (EX signal in Fig.3).

(iii) The background event is mainly due to Compton scattering of high energy *γ* rays. Since the scattered γ ray is detected by another segments, the background events is efficiently identified and rejected.

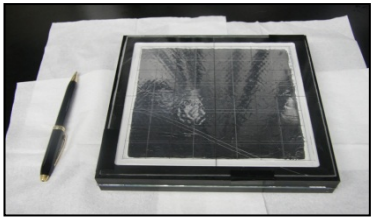

Fig.2. The thin NaI(Tl) detector PICO-LON

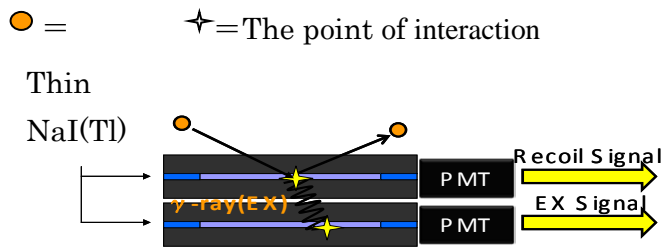

Fig.3. The segmentation system. EX stands for inelastic excitation.

Good energy resolution is required to investigate the EX type interaction because the γ rays by the EX type interaction and by background radioactivities must be distinguished. The required energy resolution for dark matter search by scintillator is better than 20% in FWHM (Full Width Half Maximum) in the low energy region [7].

## 3. EXPERIMENT

We measured the performance of three modules of NaI(Tl) detector named PICO-LON No1,

No2, and No3. The dimension of the NaI(Tl) crystal was 150mm×150mm×1mm. The good energy resolution and the low energy threshold are important to search for the dark matter. Moreover, the good position resolution is also important to identify the point of interaction [5][6][7].

The scintillation photons from NaI(Tl) are collected by 12 PMTs from thinner edges of the NaI(Tl) crystal through a light guide. The high voltages for each PMT were adjusted to give the same gain. The PMT outputs were individually connected to a charge ADC (REPIC RPC-022) and integrated the current pulse. The energy resolution and the low energy threshold were measured by using low energy γ rays and X rays from $^{241}$Am and $^{133}$Ba. The pulse height spectra for $^{241}$Am and $^{133}$Ba are measured by using a collimator made of 2cm thick lead with a 2mm $\phi$ hole which was placed above the NaI(Tl) crystal.

The position resolution in the wide area of the thin NaI(Tl) was estimated. The position resolution was measured with collimated 60keV γ ray from $^{241}$Am. The position resolution was measured at intervals of 3cm from the center. The position was shown in Fig. 4.

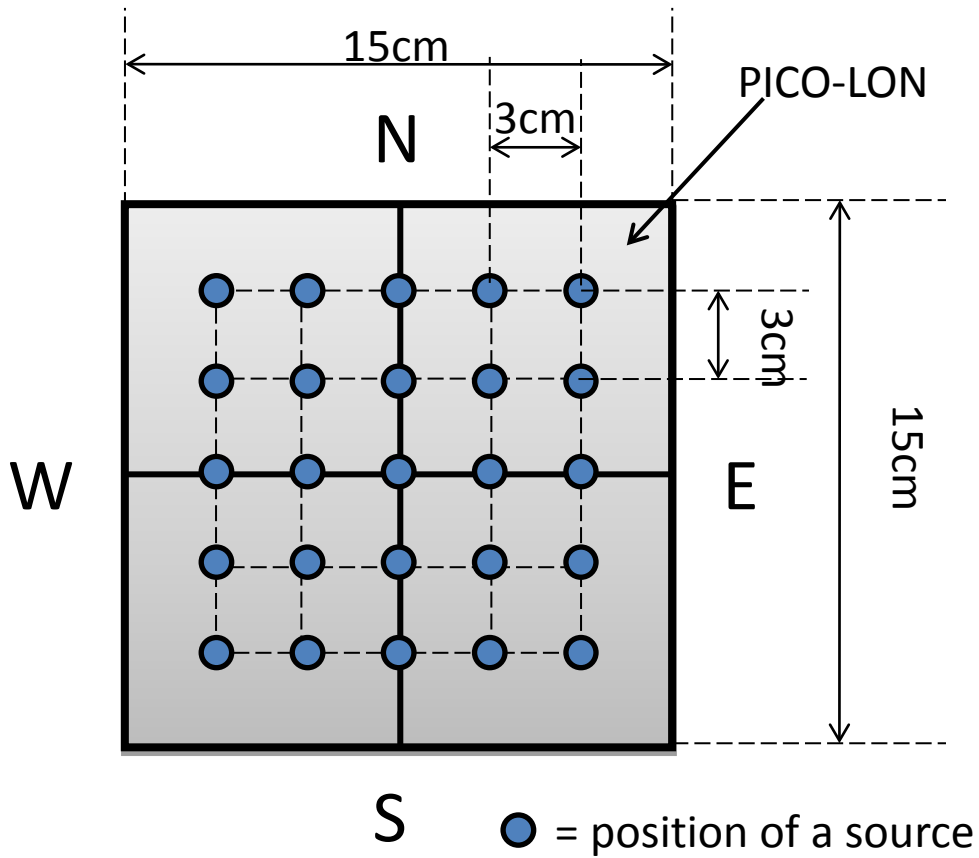

Fig.4. Position information was measured at the 25 points. The characters N, E, S and W stand for North, East, South and West.

## 4. RESULTS

The pulse height spectrum was obtained by simply summing all the ADC output. The pulse height spectrum was well reproduced by Gaussian functions and constant background. The energy resolution of each peak was evaluated from the fit.

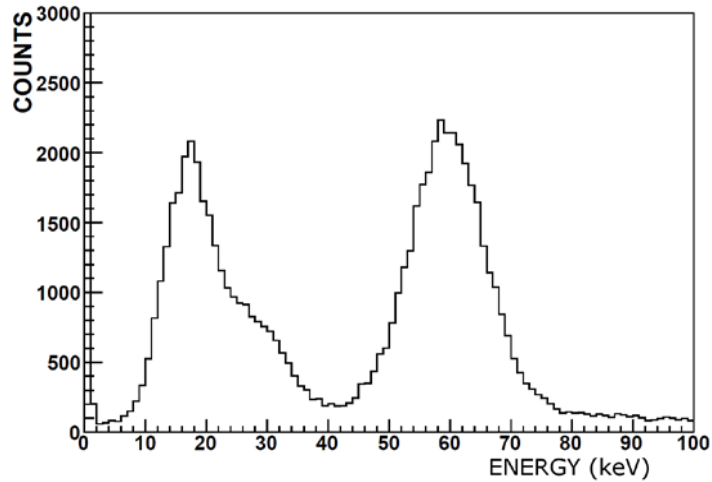

Fig.5. The energy spectrum of NaI(Tl) irradiating $^{241}$Am source.

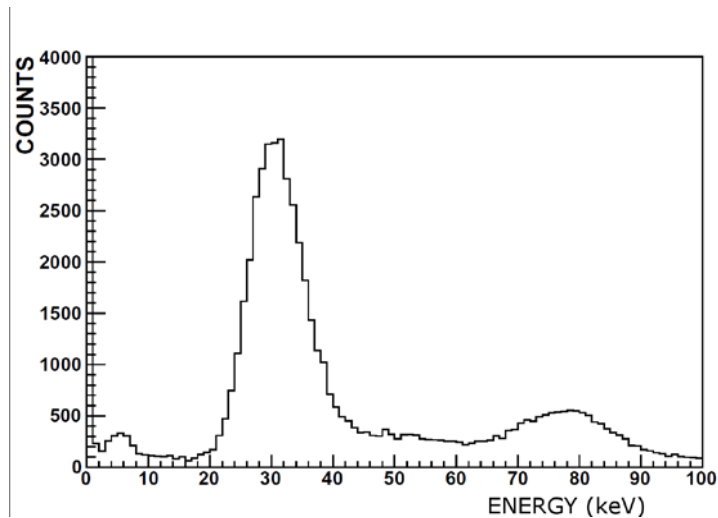

Fig.6. The energy spectrum irradiating $^{133}$Ba source.

The good energy resolution was obtained as 25% at 60keV in FWHM. The energy threshold was determined to be about 2keV. The low energy LX ray of $^{133}$Ba (4.3keV) was clearly observed as shown in Fig.6.

The position information was derived from the ratio of PMT outputs. The three PMT outputs on each side (north, east, south and west) were used for the position analysis by using following equation.

$$x = \frac{E-W}{E+W}$$
$$y = \frac{N-S}{N+S},$$

where $N, E, S$ and $W$ are the sums of three PMT outputs from northern, eastern, southern and western sides, respectively.

The result of position resolution was shown in Fig. 7. They are transformed to real scale. The position resolution of X-axis and Y-axis were 30% and 34% in FWHM. The position resolution in the wider side (15cm×15cm) was as good as 4.5cm in FWHM.

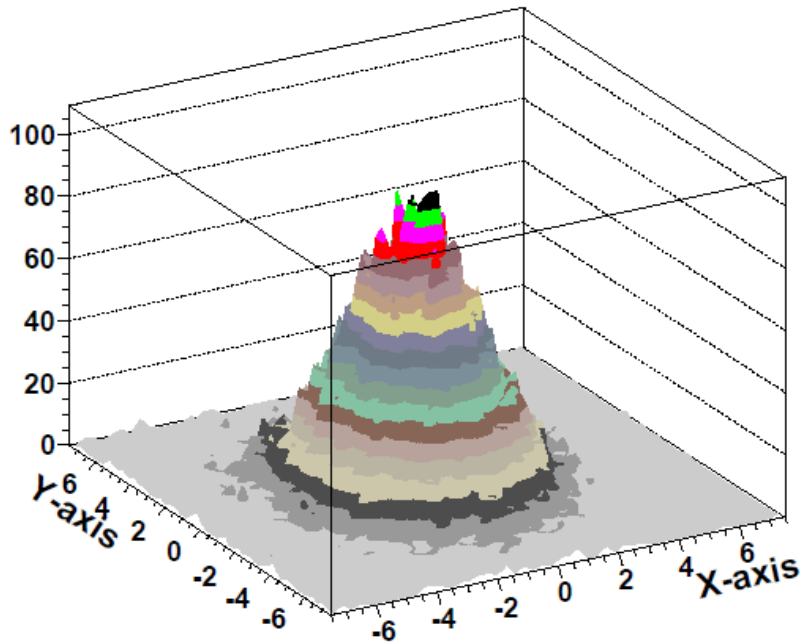

Fig.7. The two-dimensional image of the position resolution by a collimated γ ray source of $^{241}$Am.

## 5. CONCLUSION

Three thin and wide area NaI(Tl) scintillator was developed to search for the dark matter. The energy resolution in FWHM, the energy threshold and the position resolution were measured by low energy gamma rays and X rays. The energy threshold was determined to be about 2keV, which is the most important property for the dark matter search. The results are listed in Table 1.

The energy resolution and the energy threshold of thin NaI(Tl) scintillators were good enough to search for WIMPs dark matter. One module named No.2 showed worse performance than the other two modules. The reason why the performance of No.2 was worse is that the noise reduction analysis did not work effectively.

Table 1. The performance of PICO-LON.

|  | Goal | No1 | No2 | No3 |
|---|---|---|---|---|
| Energy resolution | > 25% | 25% | 26% | 24% |
| X position resolution | > 30% | 30% | 33% | 31% |
| Y position resolution | > 30% | 34% | 33% | 34% |
| Energy threshold | > 5keV | 2keV | 10keV | 2keV |

The thin NaI(Tl) has showed that it should be applied for future dark matter search. To enlarge the background reduction efficiency, an active shield with $4\pi$ solid angle will be applied. The KamLAND detector system is the good candidate to install PICO-LON system. The expected sensitivity reaches to $10^{-5} \sim 10^{-6}$ pb.

## 6. Acknowledgement

The present work was supported by JSPS Grant-in-Aid for Scientific Research (C) 20540294.